\documentclass[fleqn,10pt]{wlscirep}
\usepackage[utf8]{inputenc}
\usepackage[T1]{fontenc}

\usepackage{algorithm}
\usepackage{algpseudocode}
\usepackage{subfig}

\title{On the evolutionary cognitive pressure for experiential awareness: do machines need it?}

\author{Warisa Sritriratanarak, Paulo Garcia}
\affil{International School of Engineering, Chulalongkorn University, Bangkok, Thailand}

\affil{\{warisa.s,paulo.g\}@chula.ac.th}


\begin{abstract}
The consciousness standing for artificial intelligence divides opinions across epistemological positions.
Whether or not  machines can be conscious, and whether  we can ascertain the truth of such a proposition for any given case, has consequential ethical implications. This challenge is exacerbated by the lack of consensus on the nature of consciousness.
We address an orthogonal problem: regardless of this nature of, is it \textit{required} for machines? Specifically, we focus on a constituent element of consciousness -experiential awareness- and examine why it arose evolutionarily in biological organisms, from a computational perspective.
We show that, because of evolutionary "baggage" -autonomous neurological reactions- experiential awareness is necessary for higher-level reasoning to be possible.
The implication is that, given artificial systems are architected without such legacy considerations, it is possible to design them with an arbitrary level of intelligence, without the need for experiential awareness.
This possibility simplifies ethical considerations on artificial intelligence, and opens new approaches to the discernment of artificial consciousness.

\end{abstract}
\begin{document}

\flushbottom
\maketitle
\thispagestyle{empty}

\section{Introduction}	

Agreement on the precise nature of consciousness appears to be at an epistemological crossroads \cite{facco2017science}. Whilst theories of consciousness are far too varied to fit into a simple linear spectrum, we can nonetheless identify two diametrically opposed camps that suffice to illustrate how far apart some positions are. On one side, we have the Chalmers' Hard Problem position \cite{chalmers2017hard}, placing consciousness above the realm of scientific inquiry. On the other side, Dennett's User Illusion position \cite{dennett2016illusionism} completely explains (according to some critics, explains \textit{away} \cite{siewert1993dennett}) the mechanisms of consciousness through a physicalist perspective. The need for a globally accepted perspective on consciousness has acquired increasing ethical importance given the possible (or not) rise of artificial consciousness, with the moral implications that go with it.
\par In this paper, we adopt an approach orthogonal to the nature of consciousness: we address the problem of what it is \textit{for}, rather than what it \textit{is}. We question what it is for not through a moral, ethical, or philosophical, perspective, but from an evolutionary perspective: what does consciousness contribute to the evolutionary fitness \cite{grinde2024consciousness} of an organism? In particular, we focus on the awareness of the current experience (phenomenal or not, depending on the reader's perspective: we take no stance on this) and how it is necessary for a cognitive organism -one that perceives the outside world and can reason to achieve a goal in that world- to carry out such cognition, given its evolutionary baggage. We will show that artificial entities need not exhibit such experiential awareness: both competence and comprehension can be achieved without it. Thus, our thesis is that we can effect artificial systems of arbitrary degree of intelligence without any experiential awareness, even if one accepts artificial consciousness is possible.

\subsection{On terminology}

Because of the lack of agreement on meaning, it is challenging to use terms (\textit{consciousness, awareness, experience,} etc.) with semantic rigor without conflicting with readers' prior conceptions. We have indulged in such lax semantics in the prior section, but for the remainder of this paper we wish to be precise: thus, we will explicitly avoid the use of such overloaded words and provide a working definition for the terms we will use here. We abandon this careful use in the 'Related Work' section near the end of the paper, where we use terms as they are employed by referenced work.
\par We will use the term 'organism' to refer to a biological entity that developed through natural section, such as a mammal. We will explicitly state 'artificial systems' when referring to designed artificial intelligences, embodied or not. 'Entity' will be used to refer to both organisms and artificial systems, in general. \textit{Perception} is defined as an entity's ability to sense the outside world and translate those sensed data into symbolic internal percepts \cite{melloni2021making,frankish2007anti,nunez2019happened}, accurately or not; it makes no difference in our case whether those percepts are a correct representation of reality or not. \textit{Reasoning} is defined as the ability to use current and past percepts to predict the future, given available actions that can be performed; i.e., the ability for rational decision-making. An entity is deemed \textit{cognitive} if it is capable of both perception and reasoning, we will analyze such cognition computationally. \textit{Experiential awareness} is defined as an entity's ability to identify a \textit{here and now}: to possess a clear distinction between past and present moments, self-referencing itself in its model of the present. We do not make the claim that these aspects constitute consciousness, but we believe stating they are \textit{constituents} of consciousness, whatever its nature is, will be accepted regardless of prior position.

\section{Background}
Reasoning can be modeled in myriad ways: one of the classic formulations is that of state-space exploration \cite{zhang1999state}. Having (approximately) inferred the state of the world through an inference function post perception and knowing the effect each action at its disposal will have on that state (deterministically or probabilistically), an entity can construct a decision tree \cite{bui2020prediction} to examine the possible futures within its reach, given its reasoning horizon (how many steps in the future it can predict in useful time). If equipped with a measure of how "good" each possible future state is (usually referred to as a \textit{utility function}), the entity can then decide on which action to perform to maximize its utility (generally by choosing the action with highest expected utility). This formulation of a reasoning entity, dating back to the 1960s \cite{vanderbrug1975state}, is based on cognitive science that explored decision-making in humans, under the premise that "cognition is computation of representations" \cite{nunez2019happened}. Informally, it corresponds to "what if" scenarios \cite{edwards1954theory}; scenarios we envision to assess the likely outcome of our actions. When examining these scenarios, we are imagining sensory input (what we will see and hear; whether it will hurt or not) and imagining perception on that input, and that fictional data and percepts are constructed by accessing our memory from past sensory input and using prior information to construct hypothetical novel one \cite{huang2024neuronal,dake2025perturbing,vredeveldt2015eye}. 
\par Both biological and digital substrates are finite in size, and both are made of interconnected sub-systems that are specialized for distinct tasks \cite{mashour2020conscious,chitty2022neural}. Thus, the use of the same sub-system (e.g., the visual cortex in humans \cite{niell2021cortical,li2025neural}, graphics cards on computers \cite{chen2023interference}) by different components of the reasoning process must be multiplexed in time. When biological substrates process the recall of images from memory, the visual cortex is engaged \cite{huang2024neuronal,dake2025perturbing} (albeit with different neuronal firing patterns than those observed when processing stimuli from the senses in real time, and affecting sensing \cite{shi2024task}). Thus, all substrates must necessarily perform some resource allocation \cite{alonso2014resource}, as they are bound in \textit{structure} and \textit{function} \cite{borger2020behavioural}. When asked to recall an image as vividly as possible, most participants will close their eyes to prevent real-time sensory stimuli from interfering with the visual processing from memory in the visual cortex \cite{vredeveldt2015eye}. This implies sub-systems in our biological substrate used for real-time sensory processing (e.g., the visual cortex) are necessarily employed to process hypothetical sensory input when reasoning \cite{huang2024neuronal,dake2025perturbing,vredeveldt2015eye}.  A final property of biological substrates is relevant: sub-systems are linked to autonomous neurological responses \cite{mederos2025overwriting}, and these can be triggered at the sensing or perception level, before that percept is forwarded to higher-level reasoning \cite{hudson2000pain}, eliciting responses in, e.g., the hippocampus and amygdala, in the case of fear \cite{mobbs2009threat}. These stimulus-response autonomous properties \cite{kim2021artificial} are not surprising, considering that higher-level reasoning is much younger, in evolutionary terms, than primitive responses such as the fight-or-flight mechanism \cite{gerson2009fight}.

\section{The evolution of experiential awareness}

The first neuron emerged roughly 800 million years ago, during the Tonian period \cite{villegas2000origin}. A sufficiently complex network of neurons, arguably deserving of being called a "brain", emerged in the Cambrian explosion 200 million years later \cite{reichert2001developmental}. During this period, organisms developed the capability to distinguish external stimuli between positive and negative (i.e., valence \cite{erwin2009early}) and were thus capable of reacting to stimuli in such a way that contributed to their fitness: e.g., by avoiding negative stimuly which would cause them harm and heading toward positive stimuli that kept them fed. At this point in evolution, organisms were purely \textit{reactive}: each action would be determined solely in function of input at current time. This was the beginning of autonomous neurological responses: an action that leads to avoiding negative stimuli, if triggered immediately by such stimuli, improved the fitness of an organism (i.e., improved the chances that it would survive and reproduce, passing on such response). The mechanisms that created a causal link between stimuli and effect, leading to the association to actions, are the same that power modern reinforcement learning \cite{neftci2019reinforcement}. The mechanisms require \textit{sensing} the external world, but do not require higher-level \textit{perception} nor any form of \textit{reasoning} to be enacted; we would not likely deem such organisms cognitive.
\par By the late Triassic to middle Jurassic period, roughly 200 million years ago, mammals have developed a neocortex \cite{northcutt1995emergence}. With it, came the ability to perform \textit{simulations}: examine possible hypothetical futures, given available actions, and decide on the action that predicts the best chance of survival. Such capabilities were essential for mammals of the time to survive a world ruled by much larger creatures. Notice that this ability to "plan" was certainly not on the same timescales as human planning; rather, it enabled short-term planning, such as the optimal placement of a limb on a branch during movement towards food. We see here a primitive form of perception, where sustenance can be identified by the visual cortex and translated into symbolic information that informs cognition. This form of predictive planning is still the basis of human movement today \cite{clark2015radical} (adopted in robotics in the form of model predictive control \cite{holkar2010overview}) and is likely the evolutionary genesis of our cognitive capability today. Evolution, by nature, carries baggage; all the neuronal circuitry that enabled this mammal's ancestors to survive in the Cambrian period was still present, and remains to this day; our recoil reflex in the presence of pain being the most striking one. Evolution is also an efficient allocator of resources: once neuronal circuitry is developed for a certain purpose, it often gets re-used for similar purposes \cite{konstantinides2025neuronal}.
\par Given our definition of experiential awareness (an entity's ability to identify a \textit{here and now}: to possess a clear distinction between past and present moments, self-referencing itself in its model of the present), let us examine a Jurassic mammal and its capability to simulate the future, and see that it necessarily requires experiential awareness: that it is a necessary condition for its \textit{reasoning} to be possible. At any moment in time, the organism is experiencing present input. Its neocortex is building a simulation of possible futures; but how is this simulation being generated? Through a generative process \cite{clark2012dreaming} that is estimating future world states given past experience, given \textit{memory} of past world states. This is, of course, the evolutionary basis for memory; the adage "remembering the past to avoid repeating mistakes" is grounded in evolutionary fitness \cite{wilson2011adaptive}. In these hypothetical futures, their quality, which will eventually lead to a decision, is estimated given hypothetical stimuli (avoidance of negative inputs, seeking positive inputs).
\par At any given time, the organism's brain is entertaining \textit{past} world states, pattern-matched with \textit{present} world state to be used in conjunction for generating possible \textit{future} world states. Stimuli for the present state (real) and past and future states (recalled/imagined) is activated, and processed by the same neural circuitry. Is the organism \textit{aware} of its current experience? Does it \textit{know} (at a subconscious, computational level in its neural circuitry) what is currently real in the present moment, and what is not? It must be. For, if not, it would be autonomously reacting to recalled or hypothesized stimuli in ways that negatively affect its fitness. In other words, it is evolutionarily advantageous for a cognitive organism to have experiential awareness, as it allows it to reason more effectively. In the next section, we develop a computational model of this idea, formally showing how this distinction between real and imagined stimuli must be discerned.

\section{Computational cognition}

Let a reasoning entity be modeled by $\frac{R}{P}$, meaning a reasoning process $R$ executing on a physical substrate $P$. The physical substrate $P$ is modeled by the tuple $\langle s_t, \sigma,\pi,a^1,a^2,...,a^i \rangle$, where $s_t$ is the state of the physical substrate at time $t$, including its spatial position in the world and the values of all its internal properties; $\sigma$ is the sensing function producing data from world state $w_t$, such that $\sigma(w_t) = \{d^1,d^2,...,d^n\}$ ($d^i, i \in [1:n]$ corresponds to one datum); $\pi_t$ is the perception function at time $t$, such that $\pi_t(\sigma(w_t),KB_{1:t-1}) = \{p^1_t,p^2_t,...,p^m_t\}$ ($p^i_t, i \in [1:m]$ corresponds to one percept, and $KB_{1:t-1}$ is the collection of all percepts thus far and rules for their semantic association); and $a^1,a^2,...,a^i$ are physical actions that change the state of the outside world (one of them being the null action). $KB$ represents the entity's \textit{Knowledge Base}, i.e., the total sum of perceptual knowledge obtained thus far.
\par The reasoning process $R$ is modeled by the tuple $\langle KB_{1:t-1},\iota,\tau,d\rangle$, where $\iota$ is the world state inference function such that $w\prime_t = \iota(KB_{1:t-1},\pi_t(\sigma(w_t))), p^P_t \in w\prime_t$, where $w\prime_t$ is the approximate inference of world state and $p^P_t$ represents the entity in the world; $\tau$ is the state transition function, such that $\tau(a^j,w_t) = w^j_{t+1}, j \in [1:i]$; $\nu$ is the utility function that scores states, such that $\nu(w) = u, u \in \mathbb{R}$; and $d$ is the reasoning process, corresponding to a tree of world states connected via physical actions $a^1,a^2,...,a^i$, constructed through the successive application of $\tau$ and $\nu$ to $w\prime_t$. Notice that, while we are using a state-space exploration formulation of reasoning, this is not a unique option; other formulations (including, but not limited to, statistical inference, inductive and deductive logic, hybrid approaches) are equally valid, and do not affect the main result as long as that formulation includes a world model.

We will denote the effect of executing $f$ (where $f$ is a function or an action) by $\frac{R}{P} \xrightarrow{f} \frac{R\prime}{P\prime}$. Consider the following process of a reasoning entity, assuming two physical actions and a reasoning horizon of 2. By sensing and perceiving the world, the entity updates its KB with percepts at time $t$:

\begin{equation}
\frac{\langle KB_{1:t-1},\iota,\tau,d=\emptyset\rangle}{\langle s_t, \sigma,\pi_t,a^1,a^2 \rangle}
\xrightarrow{\pi(\sigma(w_t),KB_{1:t-1})}
\frac{\langle \{KB_{1:t-1} \cup {p^1_t,p^2_t,...,p^m_t}\},\iota,\tau,d=\emptyset\rangle}{\langle s_t, \sigma,\pi_t,a^1,a^2 \rangle}
\end{equation}

Through the  world state inference function $\iota$ applied to all percepts obtained thus far, the entity constructs an internal model of the world at time $t$:

\begin{equation}
\frac{\langle KB_{1:t-1} \cup \{p^1_t,p^2_t,...,p^m_t\},\iota,\tau,d=\emptyset\rangle}{\langle s_t, \sigma,\pi_t,a^1,a^2 \rangle}
\xrightarrow{\iota(KB_{1:t-1} \cup \{p^1_t,p^2_t,...,p^m_t\})}
\frac{\langle KB_{1:t} ,\iota,\tau,w\prime_t\rangle}{\langle s_t, \sigma,\pi_t,a^1,a^2 \rangle}
\end{equation}

By applying the transition function $\tau$ to its internal model of the world, given its available actions, the entity begins constructing a tree of future world states (i.e., its prediction of possible worlds at $t+1$):

\begin{equation}
\frac{\langle KB_{1:t} ,\iota,\tau,w\prime_t\rangle}{\langle s_t, \sigma,\pi_t,a^1,a^2 \rangle}
\xrightarrow{\tau(a^1,w\prime_t),\tau(a^2,w\prime_t)}
\frac{\left\langle KB_{1:t} ,\iota,\tau,w\prime_t \rightarrow 
    \begin{cases} 
        (a^1) , w\prime^1_{t+1} \\
        (a^2) , w\prime^2_{t+1} 
    \end{cases}\right\rangle 
}{\langle s_t, \sigma,\pi_t,a^1,a^2 \rangle}
\end{equation}

The utility function $\nu$ assigns a value to each hypothetical world state:

\begin{equation}
\frac{\left\langle KB_{1:t} ,\iota,\tau,w\prime_t \rightarrow 
    \begin{cases} 
        (a^1) , w\prime^1_{t+1} \\
        (a^2) , w\prime^2_{t+1} 
    \end{cases}\right\rangle 
}{\langle s_t, \sigma,\pi_t,a^1,a^2 \rangle}
\xrightarrow{\nu(w^1\prime_t),\nu(w^2\prime_t)}
\frac{\left\langle KB_{1:t} ,\iota,\tau,w\prime_t \rightarrow 
    \begin{cases} 
        (a^1,x_1) , w\prime^1_{t+1} \\
        (a^2,x_2) , w\prime^2_{t+1} 
    \end{cases}\right\rangle 
}{\langle s_t, \sigma,\pi_t,a^1,a^2 \rangle}
\end{equation}

assuming $x_1 = \nu(w^1\prime_t)$ and $x_2 = \nu(w^2\prime_t)$. These hypothetical world states include a representation of the entity itself in that future; i.e., $w\prime^1_{t+1} = \{w\prime^1_{t+1} \setminus p^{P_1}_{t+1},  \frac{\langle KB_{1:t},\iota,\tau,d=\emptyset\rangle}{\langle s_{t+1}, \sigma,\pi_{t+1},a^1,a^2 \rangle}\}$:

\begin{equation}
\tiny
\frac{\left\langle KB_{1:t} ,\iota,\tau,w\prime_t \rightarrow 
    \begin{cases} 
        (a^1,x_1) , w\prime^1_{t+1} \\
        (a^2,x_2) , w\prime^2_{t+1} 
    \end{cases}\right\rangle 
}{\langle s_t, \sigma,\pi_t,a^1,a^2 \rangle}
=
\frac{\left\langle KB_{1:t} ,\iota,\tau,w\prime_t \rightarrow 
    \begin{cases} 
        (a^1,x_1) , \{w\prime^1_{t+1} \setminus p^{P_1}_{t+1},  \frac{\langle KB_{1:t},\iota,\tau,d=\emptyset\rangle}{\langle s_{t+1}, \sigma,\pi_{t+1},a^1,a^2 \rangle}\} \\
        (a^2,x_2) , \{w\prime^2_{t+1} \setminus p^{P_2}_{t+1},  \frac{\langle KB_{1:t},\iota,\tau,d=\emptyset\rangle}{\langle s_{t+1}, \sigma,\pi_{t+1},a^1,a^2 \rangle}\} 
    \end{cases}\right\rangle 
}{\langle s_t, \sigma,\pi_t,a^1,a^2 \rangle}
\end{equation}

The entity computes a further step into the future by considering what it would reason about, it it were in such a hypothetical world state:

\begin{equation}
\begin{aligned}
\tiny
\frac{\left\langle KB_{1:t} ,\iota,\tau,w\prime_t \rightarrow 
    \begin{cases} 
        ...  \frac{\langle KB_{1:t},\iota,\tau,d=\emptyset\rangle}{\langle s_{t+1}, \sigma,\pi_{t+1},a^1,a^2 \rangle}\} \\
        ...  \frac{\langle KB_{1:t},\iota,\tau,d=\emptyset\rangle}{\langle s_{t+1}, \sigma,\pi_{t+1},a^1,a^2 \rangle}\} 
    \end{cases}\right\rangle 
}{\langle s_t, \sigma,\pi_t,a^1,a^2 \rangle}
\xrightarrow{\pi(\sigma(w^1\prime_{t+1}),KB_{1:t})}\\
\xrightarrow{\pi(\sigma(w^1\prime_{t+1}),KB_{1:t})}
\frac{\left\langle KB_{1:t} ,\iota,\tau,w\prime_t \rightarrow 
    \begin{cases} 
        ...  \frac{\langle \{KB_{1:t} \cup {p^1_{t+1},,...,p^m_{t+1}}\},\iota,\tau,d=\emptyset\rangle}{\langle s_{t+1}, \sigma,\pi_{t+1},a^1,a^2 \rangle}\} \\
        ...  \frac{\langle KB_{1:t},\iota,\tau,d=\emptyset\rangle}{\langle s_{t+1}, \sigma,\pi_{t+1},a^1,a^2 \rangle}\} 
    \end{cases}\right\rangle 
}{\langle s_t, \sigma,\pi_t,a^1,a^2 \rangle}
\end{aligned}
\end{equation}

finally obtaining an estimation of the world state after two actions:

\begin{equation}
\begin{aligned}
\tiny
\frac{\left\langle KB_{1:t} ,\iota,\tau,w\prime_t \rightarrow 
    \begin{cases} 
        ...  \frac{\langle \{KB_{1:t} \cup \{p^1_{t+1},,...,p^m_{t+1}\}\},\iota,\tau,d=\emptyset\rangle}{\langle s_{t+1}, \sigma,\pi_{t+1},a^1,a^2 \rangle}\} \\
        ...  \frac{\langle KB_{1:t},\iota,\tau,d=\emptyset\rangle}{\langle s_{t+1}, \sigma,\pi_{t+1},a^1,a^2 \rangle}\} 
    \end{cases}\right\rangle 
}{\langle s_t, \sigma,\pi_t,a^1,a^2 \rangle}
\xrightarrow{\iota(KB_{1:t} \cup \{p^1_{t+1},,...,p^m_{t+1}\})}\\
\xrightarrow{\iota(KB_{1:t} \cup \{p^1_{t+1},,...,p^m_{t+1}\})}
\frac{\left\langle KB_{1:t} ,\iota,\tau,w\prime_t \rightarrow 
    \begin{cases} 
        ...  \frac{\langle \{KB_{1:t+1},\iota,\tau,d=w\prime^1_{t+1}\rangle}{\langle s_{t+1}, \sigma,\pi_{t+1},a^1,a^2 \rangle}\} \\
        ...  \frac{\langle KB_{1:t},\iota,\tau,d=\emptyset\rangle}{\langle s_{t+1}, \sigma,\pi_{t+1},a^1,a^2 \rangle}\} 
    \end{cases}\right\rangle 
}{\langle s_t, \sigma,\pi_t,a^1,a^2 \rangle}
\end{aligned}
\end{equation}

where the decision process in $p^{P_1}_{t+1}$ and $p^{P_2}_{t+1}$ must be completed to reach the reasoning depth of 2 across all branches: the action that leads to the state with the highest utility would then be executed. Assuming $a_1$:

\begin{equation}
\begin{aligned}
\frac{\left\langle KB_{1:t} ,\iota,\tau,w\prime_t \rightarrow 
    \begin{cases} 
        (a^1,x_1) , \{w\prime^1_{t+1} \setminus p^{P_1}_{t+1},  \frac{\langle KB_{1:t},\iota,\tau,d=...\rangle}{\langle s_{t+1}, \sigma,\pi_{t+1},a^1,a^2 \rangle}\} \\
        (a^2,x_2) , \{w\prime^2_{t+1} \setminus p^{P_2}_{t+1},  \frac{\langle KB_{1:t},\iota,\tau,d=...\rangle}{\langle s_{t+1}, \sigma,\pi_{t+1},a^1,a^2 \rangle}\} 
    \end{cases}\right\rangle 
}{\langle s_t, \sigma,\pi_t,a^1,a^2 \rangle}
\xrightarrow{a_1}\\
\xrightarrow{a_1}
\frac{\langle KB_{1:t},\iota,\tau,d=\emptyset\rangle}{\langle s_{t+1}, \sigma,\pi_{t+1},a^1,a^2 \rangle}
\end{aligned}
\end{equation}

We are omitting the evolution of $\pi$ for now. This is the algorithm for state-space exploration utilized in classic AI agents, with additional modeling of substrate machinery. In the context of AI, this algorithm suffers from scalability issues (combinatorial explosion of hypothetical world states as the reasoning horizon increases), but performs reasonably well (within computation and memory constraints) for small reasoning horizons.
\par Let us now consider the impact of neurological autonomous reactions on this formalism. Substrate machinery is used for reasoning, and it may produce autonomous responses: let us denote an autonomous response $a_r$ triggered by an action or function $f$ by $\frac{R}{P} \xrightarrow[a_r]{f} \frac{R\prime}{P\prime}$. Let us examine Equation 6, in a scenario where perception identifies a threat in hypothetical state $w^1\prime_{t+1}$, triggering the fight or flight response, with its associated rise in stress levels, adrenaline production, etc.

\begin{equation}
\begin{aligned}
\tiny
\frac{\left\langle KB_{1:t} ,\iota,\tau,w\prime_t \rightarrow 
    \begin{cases} 
        ...  \frac{\langle KB_{1:t},\iota,\tau,d=\emptyset\rangle}{\langle s_{t+1}, \sigma,\pi_{t+1},a^1,a^2 \rangle}\} \\
        ...  \frac{\langle KB_{1:t},\iota,\tau,d=\emptyset\rangle}{\langle s_{t+1}, \sigma,\pi_{t+1},a^1,a^2 \rangle}\} 
    \end{cases}\right\rangle 
}{\langle s_t, \sigma,\pi_t,a^1,a^2 \rangle}
\xrightarrow[a_r]{\pi(\sigma(w^1\prime_{t+1}),KB_{1:t})}\\
\xrightarrow[a_r]{\pi(\sigma(w^1\prime_{t+1}),KB_{1:t})}
\frac{\left\langle KB_{1:t} ,\iota,\tau,w\prime_t \rightarrow 
    \begin{cases} 
        ...  \frac{\langle \{KB_{1:t} \cup {p^1_{t+1},,...,p^m_{t+1}}\},\iota,\tau,d=\emptyset\rangle}{\langle s_{t+1}, \sigma,\pi_{t+1},a^1,a^2 \rangle}\} \\
        ...  \frac{\langle KB_{1:t},\iota,\tau,d=\emptyset\rangle}{\langle s_{t+1}, \sigma,\pi_{t+1},a^1,a^2 \rangle}\} 
    \end{cases}\right\rangle 
}{\langle s\prime_t = a_r(s_t), \sigma,\pi_t,a^1,a^2 \rangle}
\end{aligned}
\end{equation}

This is an effect, on the real substrate state $s_t$ within $w_t$, of an imagined threat in a hypothetical state $w^1\prime_{t+1}$: damage caused by \textit{thinking}. It is straightforward to imagine many more scenarios where this damage is even move extreme, simply by considering autonomous movement (e.g., recoil as pain response) to hypothetical future states. What we have identified here is a mis-alignment between autonomous responses evolved when we were stateless reactive systems and later capability for reasoning; mis-alignment stemming from using the same substrate resources for different functions.

There is a simple solution to this mis-alignment: a switch in substrate machinery that allows \textit{turning off} (or dampening) autonomous responses; a switch (denoted $\mathbf{c} \in R$) that must necessarily be controlled by the reasoning process, such that:

\begin{equation}
    \frac{R}{P} \xrightarrow[a_r \cap \mathbf{c} = \top]{f} \frac{R\prime}{P}, \frac{R}{P} \xrightarrow[a_r \cap \mathbf{c} = \bot]{f} \frac{R\prime}{P\prime}
\end{equation}

In other words, this model allows reformulating Equation 6 such that:

\begin{equation}
\begin{aligned}
\tiny
\frac{\left\langle\mathbf{c} = \top , KB_{1:t} ,\iota,\tau,w\prime_t \rightarrow 
    \begin{cases} 
        ...  \frac{\langle KB_{1:t},\iota,\tau,d=\emptyset\rangle}{\langle s_{t+1}, \sigma,\pi_{t+1},a^1,a^2 \rangle}\} \\
        ...  \frac{\langle KB_{1:t},\iota,\tau,d=\emptyset\rangle}{\langle s_{t+1}, \sigma,\pi_{t+1},a^1,a^2 \rangle}\} 
    \end{cases}\right\rangle 
}{\langle s_t, \sigma,\pi_t,a^1,a^2 \rangle}
\xrightarrow[a_r]{\pi(\sigma(w^1\prime_{t+1}),KB_{1:t})}\\
\xrightarrow[a_r]{\pi(\sigma(w^1\prime_{t+1}),KB_{1:t})}
\frac{\left\langle\mathbf{c} = \top, KB_{1:t} ,\iota,\tau,w\prime_t \rightarrow 
    \begin{cases} 
        ...  \frac{\langle \{KB_{1:t} \cup {p^1_{t+1},,...,p^m_{t+1}}\},\iota,\tau,d=\emptyset\rangle}{\langle s\prime_{t+1} = a_r(s_{t+1}), \sigma,\pi_{t+1},a^1,a^2 \rangle}\} \\
        ...  \frac{\langle KB_{1:t},\iota,\tau,d=\emptyset\rangle}{\langle s_{t+1}, \sigma,\pi_{t+1},a^1,a^2 \rangle}\} 
    \end{cases}\right\rangle 
}{\langle s_t, \sigma,\pi_t,a^1,a^2 \rangle}
\end{aligned}
\end{equation}

In other words, an \textit{experiential awareness} switch enables such cognitive computations.

\section{Discussion}

We have seen that there is empirical evidence for the re-use of neuronal circuitry across tasks, and such re-use led evolution to have to deal with legacy features. By modeling a cognitive organism's reasoning process computationally, we have identified a mismatch, a contradiction, between these evolutionary legacy features (autonomous responses from purely reactive stages) and the novel capability to simulate hypothetical futures. Yet, we observe no such contradiction in our own reasoning process. The most plausible explanation is that evolution has very successfully dealt with such mismatch through a simple "tagging" rule, where cognitive symbols are tagged with the ability to produce autonomous responses or not. Such cognitive discrimination effects the operational functions we attribute to experiential awareness, suggesting it contributes to the fitness of a reasoning organism. More strongly, that it is a necessary condition for this level of reasoning to be possible. Thus, if we accept experiential awareness is a constituent of consciousness, we must conclude that some constituent parts are not orthogonal to intelligence, but rather its enablers. Biological organisms cannot experience intelligently without being aware of the experience.
\par What does this mean for our notions of artificial consciousness? It suggests that -even if artificial consciousness is possible- it is not necessary. Because we have the power to design the cognitive architecture of artificial systems, we are not bound by legacy evolutionary choices. We are free to implement systems where intelligence and consciousness \textit{are} orthogonal, and we can build zombie artificial systems of arbitrary intelligence.

\section{Related work}

The question of whether or not artificial systems can exhibit consciousness \cite{bengio2025illusions} is of critical importance as we consider the ethical and moral implications of ever more powerful Artificial Intelligence (AI) going forward \cite{dehaene2021consciousness}, and it addresses a fundamental question about the nature of human beings. There is no agreed-upon definition for what constitutes "consciousness", besides the general agreement that human beings exhibit it \cite{frankish2007anti}, and its definition is intertwined with several adjacent concepts, often in circular references (e.g., sentience, awareness, qualia, experience, etc.) \cite{broom2022concepts}. Beyond linguistic ambiguity, there is also an overlap between perceptions of consciousness and other concepts (e.g., wakefulness, intelligence) \cite{duch2007computational}, and cultural and religious overhead that contributes to ambiguity \cite{minsky1992conversation}. 

\par This inexact definition has challenged the design of consciousness tests \cite{jo2025quest} and, sometimes implicitly, prevents meaningful comparisons of different theories of consciousness \cite{kuhn2024landscape}. Broadly, consciousness has been divided into phenomenal and cognitive \cite{block2007consciousness}, and its study approached from the perspective of natural dualism (Chalmers' "hard problem", positing consciousness is beyond the reach of contemporary science \cite{melloni2021making}) and from the perspective of physicalism (sometimes referred to as materialism, and more formally as computational functionalism, which posits consciousness is ultimately a computational process that can be modeled and understood \cite{butlin2023consciousness}). One the main challenges is the conflation of consciousness and other cognitive functions (namely, intelligence), a topic explored in detail by Minsky \cite{minsky1992conversation,minsky2011interior,minsky1980decentralized}; we point interested readers to the extensive review by Kuhn \cite{kuhn2024landscape} for a much more in-depth description and to the work of Butlin et al \cite{butlin2023consciousness} for a clear distinction between intelligence and consciousness in the context of AI, which we adopt here.
\par Predictive perception \cite{friston2012prediction} describes the human brain as an inference machine, showing how generative world models are combined with sensory information to estimate the state of the outside world. The theory opposes the traditional view of perception as a purely feed-forward mechanism \cite{drayson2018direct}, and instead exposes the intricate feedback system that is involved in perception and, ultimately, human experience \cite{seth2019unconscious}. Predictive processing has now garnered sufficient empirical evidence \cite{walsh2020evaluating} that several syndromes can be explained by it. Recently, how inference plays a role in action decision and performance has illuminated further mechanisms \cite{knoblich2001predicting}. The discovery of predictive perception, and how it impacts  subjective experience and actions, has obviously been identified as having significant consequences for our understanding of cognition \cite{nave2022expecting,picard2014predictions,marvan2021predictive,hohwy2020predictive}; in this paper, we take this idea one step further and formalize a specific mechanism through which this occurs.
\par The existence of neuronal circuitry for distinguishing the present from hypothetical futures is supported by empirical data on sub-system neuronal activity in reality versus recall\cite{shi2024task}. The "evidence for a temporal circuit characterized by a set of trajectories along which dynamic brain activity occurs" discovered by Huang et al \cite{huang2020temporal}, regulating the activity of two distinct cortical systems (the default mode network and the dorsal attention network) is consistent with the hypothesis, suggesting different substrate sub-systems for different aspects of reasoning. Similarly, Dehaene, Lau and Kouider \cite{dehaene2021consciousness} suggest that there are "... two different types of information-processing computations in the brain: the selection of information for global broadcasting... , and the self-monitoring of those computations,...": we suggest this self-monitoring forms the basis of experiential awareness, whilst the selection of information for broadcasting corresponds to the distribution of software across substrate resources. Findings of similar neurological markers in non-human animals that we consider highly intelligence (e.g., corvid birds \cite{nieder2020neural}) support this link.

\section{Conclusions}

The main objective of this paper was to illustrate how, despite the lack of accepted agreement on a definition of consciousness, a feature accepted as a constituent part is not orthogonal to intelligence, for evolved biological organisms such as us. This link between intelligence and consciousness features must be better understood, as it has made research fuzzy in the past. We hope to help improve this discernment.
\par A conclusion that arose from this discernment is that the question of whether or not machines can \textit{be} conscious is perhaps premature; a more immediate question is \textit{need} they be? We argue 'no': our results suggest we can build artificial systems that can effect all our cognitive requirements without experiential awareness, and this view likely simplifies the ethical standing of artificial intelligences in our society.

\bibliography{sample}

@article{friston2012prediction,
  title={Prediction, perception and agency},
  author={Friston, Karl},
  journal={International Journal of Psychophysiology},
  volume={83},
  number={2},
  pages={248--252},
  year={2012},
  publisher={Elsevier}
}

@article{siewert1993dennett,
  title={What Dennett can't imagine and why},
  author={Siewert, Charles},
  journal={Inquiry},
  volume={36},
  number={1-2},
  pages={93--112},
  year={1993},
  publisher={Taylor \& Francis}
}

@article{dennett2016illusionism,
  title={Illusionism as the obvious default theory of consciousness},
  author={Dennett, Daniel C},
  journal={Journal of Consciousness Studies},
  volume={23},
  number={11-12},
  pages={65--72},
  year={2016},
  publisher={Imprint Academic}
}

@article{chalmers2017hard,
  title={The hard problem of consciousness},
  author={Chalmers, David},
  journal={The Blackwell companion to consciousness},
  pages={32--42},
  year={2017},
  publisher={Wiley Online Library}
}

@article{facco2017science,
  title={On the science of consciousness: Epistemological reflections and clinical implications},
  author={Facco, Enrico and Lucangeli, Daniela and Tressoldi, Patrizio},
  journal={Explore},
  volume={13},
  number={3},
  pages={163--180},
  year={2017},
  publisher={Elsevier}
}

@article{clark2012dreaming,
  title={Dreaming the whole cat: Generative models, predictive processing, and the enactivist conception of perceptual experience},
  author={Clark, Andy},
  journal={Mind},
  volume={121},
  number={483},
  pages={753--771},
  year={2012},
  publisher={Mind Association}
}

@article{wilson2011adaptive,
  title={Adaptive memory: Fitness relevant stimuli show a memory advantage in a game of pelmanism},
  author={Wilson, Stuart and Darling, Stephen and Sykes, Jonathan},
  journal={Psychonomic bulletin \& review},
  volume={18},
  number={4},
  pages={781--786},
  year={2011},
  publisher={Springer}
}

@article{konstantinides2025neuronal,
  title={Neuronal circuit evolution: From development to structure and adaptive significance},
  author={Konstantinides, Nikolaos and Desplan, Claude},
  journal={Cold Spring Harbor perspectives in biology},
  volume={17},
  number={5},
  pages={a041493},
  year={2025},
  publisher={Cold Spring Harbor Lab}
}

@article{clark2015radical,
  title={Radical predictive processing},
  author={Clark, Andy},
  journal={The Southern Journal of Philosophy},
  volume={53},
  pages={3--27},
  year={2015},
  publisher={Wiley Online Library}
}

@article{neftci2019reinforcement,
  title={Reinforcement learning in artificial and biological systems},
  author={Neftci, Emre O and Averbeck, Bruno B},
  journal={Nature Machine Intelligence},
  volume={1},
  number={3},
  pages={133--143},
  year={2019},
  publisher={Nature Publishing Group UK London}
}

@article{holkar2010overview,
  title={An overview of model predictive control},
  author={Holkar, Kailas S and Waghmare, Laxman M},
  journal={International Journal of control and automation},
  volume={3},
  number={4},
  pages={47--63},
  year={2010}
}

@article{northcutt1995emergence,
  title={The emergence and evolution of mammalian neocortex},
  author={Northcutt, R Glenn and Kaas, Jon H},
  journal={Trends in neurosciences},
  volume={18},
  number={9},
  pages={373--379},
  year={1995},
  publisher={Elsevier}
}

@article{erwin2009early,
  title={Early origin of the bilaterian developmental toolkit},
  author={Erwin, Douglas H},
  journal={Philosophical Transactions of the Royal Society B: Biological Sciences},
  volume={364},
  number={1527},
  pages={2253--2261},
  year={2009},
  publisher={The Royal Society}
}

@article{reichert2001developmental,
  title={Developmental genetic evidence for a monophyletic origin of the bilaterian brain},
  author={Reichert, Heinrich and Simeone, Antonio},
  journal={Philosophical Transactions of the Royal Society of London. Series B: Biological Sciences},
  volume={356},
  number={1414},
  pages={1533--1544},
  year={2001},
  publisher={The Royal Society}
}

@inproceedings{villegas2000origin,
  title={The origin of the neuron: The first neuron in the phylogenetic tree of life},
  author={Villegas, Raimundo and Castillo, Cecilia and Villegas, Gloria M},
  booktitle={Astrobiology: Origins from the Big-Bang to Civilisation Proceedings of the Iberoamerican School of Astrobiology Caracas, Venezuela, 28 November--8 December, 1999},
  pages={195--211},
  year={2000},
  organization={Springer}
}

@article{hohwy2020predictive,
  title={Predictive processing as a systematic basis for identifying the neural correlates of consciousness},
  author={Hohwy, Jakob and Seth, Anil},
  journal={PhiMiSci: Philosophy and the Mind Sciences},
  volume={1},
  number={2},
  pages={3},
  year={2020},
  publisher={Freie Universit{\"a}t Berlin}
}

@article{marvan2021predictive,
  title={Is predictive processing a theory of perceptual consciousness?},
  author={Marvan, Tom{\'a}{\v{s}} and Havl{\'\i}k, Marek},
  journal={New Ideas in Psychology},
  volume={61},
  pages={100837},
  year={2021},
  publisher={Elsevier}
}

@article{picard2014predictions,
  title={Predictions, perception, and a sense of self},
  author={Picard, Fabienne and Friston, Karl},
  journal={Neurology},
  volume={83},
  number={12},
  pages={1112--1118},
  year={2014},
  publisher={Lippincott Williams \& Wilkins Hagerstown, MD}
}

@article{nave2022expecting,
  title={Expecting some action: Predictive processing and the construction of conscious experience},
  author={Nave, Kathryn and Deane, George and Miller, Mark and Clark, Andy},
  journal={Review of Philosophy and Psychology},
  volume={13},
  number={4},
  pages={1019--1037},
  year={2022},
  publisher={Springer}
}

@article{knoblich2001predicting,
  title={Predicting the effects of actions: Interactions of perception and action},
  author={Knoblich, G{\"u}nther and Flach, R{\"u}diger},
  journal={Psychological science},
  volume={12},
  number={6},
  pages={467--472},
  year={2001},
  publisher={SAGE Publications Sage CA: Los Angeles, CA}
}

@article{drayson2018direct,
  title={Direct perception and the predictive mind},
  author={Drayson, Zoe},
  journal={Philosophical Studies},
  volume={175},
  number={12},
  pages={3145--3164},
  year={2018},
  publisher={Springer}
}

@article{walsh2020evaluating,
  title={Evaluating the neurophysiological evidence for predictive processing as a model of perception},
  author={Walsh, Kevin S and McGovern, David P and Clark, Andy and O'Connell, Redmond G},
  journal={Annals of the new York Academy of Sciences},
  volume={1464},
  number={1},
  pages={242--268},
  year={2020},
  publisher={Wiley Online Library}
}

@article{seth2019unconscious,
  title={From unconscious inference to the beholder’s share: Predictive perception and human experience},
  author={Seth, Anil K},
  journal={European Review},
  volume={27},
  number={3},
  pages={378--410},
  year={2019},
  publisher={Cambridge University Press}
}

@article{minsky1980decentralized,
  title={Decentralized minds},
  author={Minsky, Marvin},
  journal={Behavioral and Brain Sciences},
  volume={3},
  number={3},
  pages={439--440},
  year={1980},
  publisher={Cambridge University Press}
}

@article{minsky1992conversation,
  title={A conversation with Marvin Minsky},
  author={Minsky, Marvin L and Laske, Otto},
  journal={AI Magazine},
  volume={13},
  number={3},
  pages={31--31},
  year={1992}
}

@incollection{minsky2011interior,
  title={Interior grounding, reflection, and self-consciousness},
  author={Minsky, Marvin},
  booktitle={Information and Computation: Essays on Scientific and Philosophical Understanding of Foundations of Information and Computation},
  pages={287--305},
  year={2011},
  publisher={World Scientific}
}

@article{bui2020prediction,
  title={Prediction of slope failure in open-pit mines using a novel hybrid artificial intelligence model based on decision tree and evolution algorithm},
  author={Bui, Xuan-Nam and Nguyen, Hoang and Choi, Yosoon and Nguyen-Thoi, Trung and Zhou, Jian and Dou, Jie},
  journal={Scientific reports},
  volume={10},
  number={1},
  pages={9939},
  year={2020},
  publisher={Nature Publishing Group UK London}
}

@book{zhang1999state,
  title={State-space search: Algorithms, complexity, extensions, and applications},
  author={Zhang, Weixiong},
  year={1999},
  publisher={Springer Science \& Business Media}
}

@article{borger2020behavioural,
  title={A behavioural theory of recursive algorithms},
  author={B{\"o}rger, Egon and Schewe, Klaus-Dieter},
  journal={Fundamenta Informaticae},
  volume={177},
  number={1},
  pages={1--37},
  year={2020},
  publisher={SAGE Publications Sage UK: London, England}
}

@article{alonso2014resource,
  title={Resource allocation in the brain},
  author={Alonso, Ricardo and Brocas, Isabelle and Carrillo, Juan D},
  journal={Review of Economic Studies},
  volume={81},
  number={2},
  pages={501--534},
  year={2014},
  publisher={Oxford University Press}
}

@incollection{duch2007computational,
  title={What is Computational Intelligence and where is it going?},
  author={Duch, Wlodzislaw},
  booktitle={Challenges for computational intelligence},
  pages={1-13},
  year={2007},
  publisher={Springer}
}

@article{melloni2021making,
  title={Making the hard problem of consciousness easier},
  author={Melloni, Lucia and Mudrik, Liad and Pitts, Michael and Koch, Christof},
  journal={Science},
  volume={372},
  number={6545},
  pages={911--912},
  year={2021},
  publisher={American Association for the Advancement of Science}
}

@article{nieder2020neural,
  title={A neural correlate of sensory consciousness in a corvid bird},
  author={Nieder, Andreas and Wagener, Lysann and Rinnert, Paul},
  journal={Science},
  volume={369},
  number={6511},
  pages={1626--1629},
  year={2020},
  publisher={American Association for the Advancement of Science}
}

@article{dehaene2021consciousness,
  title={What is consciousness, and could machines have it?},
  author={Dehaene, Stanislas and Lau, Hakwan and Kouider, Sid},
  journal={Robotics, AI, and humanity: Science, ethics, and policy},
  pages={43--56},
  year={2021},
  publisher={Springer International Publishing Cham}
}

@article{huang2020temporal,
  title={Temporal circuit of macroscale dynamic brain activity supports human consciousness},
  author={Huang, Zirui and Zhang, Jun and Wu, Jinsong and Mashour, George A and Hudetz, Anthony G},
  journal={Science advances},
  volume={6},
  number={11},
  pages={eaaz0087},
  year={2020},
  publisher={American Association for the Advancement of Science}
}

@article{kim2021artificial,
  title={Artificial stimulus-response system capable of conscious response},
  author={Kim, Seongchan and Roe, Dong Gue and Choi, Yoon Young and Woo, Hwije and Park, Joongpill and Lee, Jong Ik and Choi, Yongsuk and Jo, Sae Byeok and Kang, Moon Sung and Song, Young Jae and others},
  journal={Science Advances},
  volume={7},
  number={15},
  pages={eabe3996},
  year={2021},
  publisher={American Association for the Advancement of Science}
}

@article{frankish2007anti,
  title={The Anti-Zombie Argument},
  author={Frankish, Keith},
  journal={The Philosophical Quarterly},
  volume={57},
  number={229},
  pages={650--666},
  year={2007},
  publisher={Blackwell Publishers Ltd Oxford, UK and Boston, USA}
}

@article{bengio2025illusions,
  title={Illusions of AI consciousness},
  author={Bengio, Yoshua and Elmoznino, Eric},
  journal={Science},
  volume={389},
  number={6765},
  pages={1090--1091},
  year={2025},
  publisher={American Association for the Advancement of Science}
}

@article{nunez2019happened,
  title={What happened to cognitive science?},
  author={N{\'u}{\~n}ez, Rafael and Allen, Michael and Gao, Richard and Miller Rigoli, Carson and Relaford-Doyle, Josephine and Semenuks, Arturs},
  journal={Nature human behaviour},
  volume={3},
  number={8},
  pages={782--791},
  year={2019},
  publisher={Nature Publishing Group UK London}
}

@article{edwards1954theory,
  title={The theory of decision making.},
  author={Edwards, Ward},
  journal={Psychological bulletin},
  volume={51},
  number={4},
  pages={380},
  year={1954},
  publisher={American Psychological Association}
}

@article{vanderbrug1975state,
  title={State-space problem-reduction, and theorem proving-some relationships},
  author={VanderBrug, Gordon J and Minker, Jack},
  journal={Communications of the ACM},
  volume={18},
  number={2},
  pages={107--115},
  year={1975},
  publisher={ACM New York, NY, USA}
}

@article{gerson2009fight,
  title={Of fight and flight},
  author={Gerson, Myron C and Abdul-Waheed, Mohammad and Millard, Ronald W},
  journal={Journal of nuclear cardiology},
  volume={16},
  number={2},
  pages={176--179},
  year={2009},
  publisher={Springer}
}

@article{mobbs2009threat,
  title={From threat to fear: the neural organization of defensive fear systems in humans},
  author={Mobbs, Dean and Marchant, Jennifer L and Hassabis, Demis and Seymour, Ben and Tan, Geoffrey and Gray, Marcus and Petrovic, Predrag and Dolan, Raymond J and Frith, Christopher D},
  journal={Journal of Neuroscience},
  volume={29},
  number={39},
  pages={12236--12243},
  year={2009},
  publisher={Society for Neuroscience}
}

@article{vredeveldt2015eye,
  title={Eye remember what happened: Eye-closure improves recall of events but not face recognition},
  author={Vredeveldt, Annelies and Tredoux, Colin G and Kempen, Kate and Nortje, Alicia},
  journal={Applied Cognitive Psychology},
  volume={29},
  number={2},
  pages={169--180},
  year={2015},
  publisher={Wiley Online Library}
}

@article{hudson2000pain,
  title={Pain perception and response: central nervous system mechanisms},
  author={Hudson, Arthur J},
  journal={Canadian journal of neurological sciences},
  volume={27},
  number={1},
  pages={2--16},
  year={2000},
  publisher={Cambridge University Press}
}

@article{li2025neural,
  title={Neural mechanisms of resource allocation in working memory},
  author={Li, Hsin-Hung and Sprague, Thomas C and Yoo, Aspen H and Ma, Wei Ji and Curtis, Clayton E},
  journal={Science Advances},
  volume={11},
  number={15},
  pages={eadr8015},
  year={2025},
  publisher={American Association for the Advancement of Science}
}

@article{dake2025perturbing,
  title={Perturbing human V1 degrades the fidelity of visual working memory},
  author={Dake, Mrugank and Curtis, Clayton E},
  journal={Nature communications},
  volume={16},
  number={1},
  pages={1--8},
  year={2025},
  publisher={Nature Publishing Group}
}

@article{niell2021cortical,
  title={How cortical circuits implement cortical computations: mouse visual cortex as a model},
  author={Niell, Cristopher M and Scanziani, Massimo},
  journal={Annual Review of Neuroscience},
  volume={44},
  number={1},
  pages={517--546},
  year={2021},
  publisher={Annual Reviews}
}

@article{shi2024task,
  title={Task-irrelevant features in working memory alter current visual processing},
  author={Shi, Hongqiao and Zhang, Qiqi and Zhou, Jieyudong and Ding, Yue and Wang, Yonghui and Li, Ya},
  journal={bioRxiv},
  pages={2024--10},
  year={2024},
  publisher={Cold Spring Harbor Laboratory}
}

@article{huang2024neuronal,
  title={Neuronal representation of visual working memory content in the primate primary visual cortex},
  author={Huang, Jiancao and Wang, Tian and Dai, Weifeng and Li, Yang and Yang, Yi and Zhang, Yange and Wu, Yujie and Zhou, Tingting and Xing, Dajun},
  journal={Science advances},
  volume={10},
  number={24},
  pages={eadk3953},
  year={2024},
  publisher={American Association for the Advancement of Science}
}

@article{mederos2025overwriting,
  title={Overwriting an instinct: Visual cortex instructs learning to suppress fear responses},
  author={Mederos, Sara and Blakely, Patty and Vissers, Nicole and Clopath, Claudia and Hofer, Sonja B},
  journal={Science},
  volume={387},
  number={6734},
  pages={682--688},
  year={2025},
  publisher={American Association for the Advancement of Science}
}

@inproceedings{chen2023interference,
  title={Interference-aware multiplexing for deep learning in gpu clusters: A middleware approach},
  author={Chen, Wenyan and Mo, Zizhao and Xu, Huanle and Ye, Kejiang and Xu, Chengzhong},
  booktitle={Proceedings of the International Conference for High Performance Computing, Networking, Storage and Analysis},
  pages={1--15},
  year={2023}
}

@article{butlin2023consciousness,
  title={Consciousness in artificial intelligence: insights from the science of consciousness},
  author={Butlin, Patrick and Long, Robert and Elmoznino, Eric and Bengio, Yoshua and Birch, Jonathan and Constant, Axel and Deane, George and Fleming, Stephen M and Frith, Chris and Ji, Xu and others},
  journal={arXiv preprint arXiv:2308.08708},
  year={2023}
}

@article{chitty2022neural,
  title={Neural architecture search survey: A hardware perspective},
  author={Chitty-Venkata, Krishna Teja and Somani, Arun K},
  journal={ACM Computing Surveys},
  volume={55},
  number={4},
  pages={1--36},
  year={2022},
  publisher={ACM New York, NY}
}

@article{mashour2020conscious,
  title={Conscious processing and the global neuronal workspace hypothesis},
  author={Mashour, George A and Roelfsema, Pieter and Changeux, Jean-Pierre and Dehaene, Stanislas},
  journal={Neuron},
  volume={105},
  number={5},
  pages={776--798},
  year={2020},
  publisher={Elsevier}
}

@article{block2007consciousness,
  title={Consciousness, accessibility, and the mesh between psychology and neuroscience},
  author={Block, Ned},
  journal={Behavioral and brain sciences},
  volume={30},
  number={5-6},
  pages={481--499},
  year={2007},
  publisher={Cambridge University Press}
}

@article{kuhn2024landscape,
  title={A landscape of consciousness: Toward a taxonomy of explanations and implications},
  author={Kuhn, Robert Lawrence},
  journal={Progress in biophysics and molecular biology},
  volume={190},
  pages={28--169},
  year={2024},
  publisher={Elsevier}
}

@article{broom2022concepts,
  title={Concepts and interrelationships of awareness, consciousness, sentience, and welfare},
  author={Broom, Donald M},
  journal={Journal of Consciousness Studies},
  volume={29},
  number={3-4},
  pages={129--149},
  year={2022},
  publisher={Imprint Academic}
}

@article{grinde2024consciousness,
  title={Consciousness makes sense in the light of evolution},
  author={Grinde, Bj{\o}rn},
  journal={Neuroscience \& Biobehavioral Reviews},
  volume={164},
  pages={105824},
  year={2024},
  publisher={Elsevier}
}

@article{jo2025quest,
  title={THE QUEST TO TEST FOR CONSCIOUSNESS},
  author={JO, AWE{\L}},
  journal={Nature},
  volume={643},
  pages={31},
  year={2025}
}

\end{document}